\documentclass{article}
\usepackage{graphicx} 
\usepackage{float}
\usepackage[utf8]{inputenc}
\usepackage[english]{babel}
\usepackage{csquotes}
\usepackage{amsmath}

\usepackage[
backend=biber,
style=alphabetic,
sorting=ynt
]{biblatex}

\title{\textbf{Hacia una implementación ética e inclusiva de la Inteligencia Artificial en las organizaciones: un marco multidimensional}}
\author{Ernesto Giralt Hernández \\ \textit{ernesto.giralt@aquarelleai.com} }

\date{Barcelona, Abril 2024}

\begin{document}
\maketitle

\begin{abstract}
El artículo analiza el impacto de la inteligencia artificial (IA) en la sociedad contemporánea y la importancia de adoptar un enfoque ético en su desarrollo e implementación dentro de las organizaciones. Se examina la perspectiva crítica del filósofo francés Éric Sadin y otros, que advierten sobre los riesgos de una tecnologización desmedida que pueda erosionar la autonomía humana. Sin embargo, el artículo también reconoce el papel activo que pueden desempeñar diversos actores, como gobiernos, académicos y sociedad civil, en dar forma al desarrollo de la IA alineado con valores humanos y sociales.

Se propone un enfoque multidimensional que combina la ética con la regulación, la innovación y la educación. Se resalta la importancia de desarrollar marcos éticos detallados, incorporar la ética en la formación de profesionales, realizar auditorías de impacto ético y fomentar la participación de las partes interesadas en el diseño de la IA.

Además, se presentan cuatro pilares fundamentales para la implementación ética de la IA en las organizaciones: 1) Valores integrados, 2) Confianza y transparencia, 3) Potenciar el crecimiento humano, y 4) Identificar factores estratégicos. Estos pilares abarcan aspectos como la alineación con la identidad ética de la empresa, la gobernanza y rendición de cuentas, el diseño centrado en el humano, la capacitación continua y la adaptabilidad frente a los cambios tecnológicos y de mercado.

Se concluye enfatizando que la ética debe ser la piedra angular en la estrategia de cualquier organización que aspire a incorporar la IA, estableciendo un marco sólido que garantice que la tecnología se desarrolle y utilice de manera que respete y promueva los valores humanos.

\end{abstract}

 \newpage
\tableofcontents
\newpage

\section{Tecnocrítica y actores claves en la era de la IA}
En los últimos años, ha surgido un discurso crítico que cuestiona el impacto de las tecnologías digitales y la inteligencia artificial (IA) en la sociedad contemporánea. Estos análisis desafían la narrativa predominante de que toda innovación tecnológica es inherentemente positiva y benéfica. En su lugar, destacan los riesgos potenciales de una tecnologización excesiva que pueda erosionar la autonomía humana e incluso tener un impacto negativo de magnitudes civilizatorias.

Este enfoque crítico, extremo pero válido, examina cómo la digitalización y el auge de la IA generativa están reconfigurando radicalmente nuestras vidas, la estructura económica y el panorama cultural. Conceptos como la "silicolonización" han sido acuñados para referirse a la imposición omnipresente de una lógica tecnológica que se filtra en todos los ámbitos de la existencia humana \cite{esadin2020}

Si bien sería inapropiado encasillar estos análisis dentro de corrientes puramente luditas — que recuerdan la revuelta contra la tecnología de la primera mitad del siglo XIX — o tecnófobas, es más acertado situarlos dentro de un tecnocriticismo matizado matizado por conceptos como el fin de la 'destrucción creativa' de Schumpeter, que rechaza la fusión indiscriminada entre el ser humano y la tecnología bajo una mirada escéptica. Este enfoque se dedica principalmente a analizar las tendencias dominantes y los actores influyentes que dan forma al desarrollo e implementación de la tecnología, con especial atención al papel de las grandes corporaciones tecnológicas.

La mirada de éstos pensadores se dedica principalmente a analizar las tendencias dominantes y los actores influyentes que configuran el desarrollo y la implementación de la tecnología, poniendo especial atención en el rol de las grandes corporaciones tecnológicas. Este enfoque revela una de las principales limitaciones de tales argumentos: si se aceptan sin un examen crítico adicional, podríamos concluir erróneamente que los actores clave del ecosistema tecnológico juegan un rol pasivo o simplemente reactivo en este proceso

Y es que la realidad puede ser más compleja y dinámica. Estudios empíricos demuestran que las corporaciones han sido actores muy decisivos, no meramente reactivos, en impulsar determinados modelos de innovación tecnológica acordes a sus intereses comerciales \cite{jobinai2019}. Y aunque es innegable que las grandes corporaciones ejercen una influencia considerable en la trayectoria del desarrollo y aplicación de la IA, es fundamental reconocer el papel igualmente primario que desempeñan estos otros actores como los gobiernos, la comunidad científica y la sociedad civil organizada. 

Hablemos de la probable influencia o rol que juega cada uno:
\begin{itemize}
    \item \textbf{Gobiernos}: A través de la regulación y la legislación, los gobiernos pueden y están intentando dar forma al desarrollo de la IA. Esto incluye todo, desde regulaciones sobre privacidad de datos hasta normativas específicas sobre el uso de IA en sectores como salud y transporte.
    \item \textbf{Científicos y académicos}:  El mundo académico contribuye no solo a través de avances tecnológicos, sino también mediante la ética en IA, estudios de impacto social y el planteamiento de alternativas al status quo tecnológico.
    \item \textbf{Sociedad civil}: Los grupos de defensa y los individuos tienen el poder de influir en el desarrollo de la tecnología mediante campañas de concienciación, presión política y la demanda de estándares éticos.
    \item \textbf{Empresas tecnológicas}: Aunque muchas grandes empresas pueden impulsar una agenda tecnológica específica, también existen numerosas startups y empresas que buscan desarrollar y promover enfoques de IA más éticos y centrados en el ser humano.
\end{itemize}

Por lo tanto, aunque se pueden señalar importantes riesgos y tendencias, también es esencial reconocer y potenciar el papel activo que pueden y deben tener todos los actores en el ecosistema de la IA para asegurar un desarrollo tecnológico que esté alineado con los valores humanos y sociales. 

\section{Cómo pueden las organizaciones participar}
El enfoque de enfatizar la creación y adopción de \textit{valores digitales} para guiar el desarrollo y uso de la inteligencia artificial es importante y ha de ser proactivo. Desde esta posición se puede reconocer que lo mismo que la tecnología no es inevitable o destructiva \textit{per se}, tampoco es neutral y que la forma en que se desarrolla, implementa y utiliza puede – y debe – alinearse con principios éticos explícitos que reflejen valores humanos universales. 

\paragraph{DIMENSIONES CLAVES}\mbox{} \\

Las principales dimensiones de cómo esta visión puede implementarse desde la mirada efectiva de una organización pueden ser:

\begin{itemize}
    \item \textbf{Desarrollo de marcos éticos}: Instituciones académicas, organismos reguladores y líderes de la industria pueden colaborar en la creación de marcos éticos detallados para la IA. Estos marcos incluirían principios como la transparencia, la justicia, la no discriminación y la responsabilidad. Ejemplos de estos esfuerzos incluyen las directrices éticas para la IA confiable de la Unión Europea. \cite{comisioneuropea2019}
    \item \textbf{Educación y formación en ética para tecnólogos}: Incorporar la ética en la educación de los ingenieros de software, científicos de datos, gestores de proyectos y líderes técnicos es vital. Esto puede ayudar a que los futuros profesionales sean conscientes de las implicaciones de su trabajo y les dotar de herramientas para tomar decisiones tecnológicas informadas por valores éticos.
    \item \textbf{Auditorías y evaluaciones de impacto ético}: Antes de implementar nuevos sistemas de IA, realizar auditorías de impacto ético podría ser obligatorio. Esto ayudaría a identificar y mitigar posibles efectos negativos en la sociedad o en grupos específicos, asegurando que los sistemas de IA se alineen con los valores sociales deseados.
    \item \textbf{Incorporación de la perspectiva de los usuarios y las partes afectadas}: Diseñar procesos participativos donde las comunidades afectadas por las tecnologías de IA puedan aportar su visión y sus preocupaciones. Esto podría ayudar a garantizar que los sistemas de IA sean inclusivos y representativos de las necesidades y valores de una gama más amplia de la sociedad.
    \item \textbf{Certificaciones y estándares para la IA ética}: Desarrollar estándares y certificaciones que puedan demostrar el cumplimiento de las normas éticas en productos y servicios de IA. Esto no solo aumentaría la confianza en estos sistemas, sino que también podría incentivar a las empresas a considerar la ética como un factor esencial en el diseño de IA.
    \item \textbf{Equipos multidisciplanarios}: Algunos de los nuevos retos de estas tecnologías, principalmente los modelos generativos, tienen que ver con la necesidad de obtener una apreciación multi-profesional: no solo necesitamos especialistas en software, datos y ciencias, sino también se requerirán lingüistas, periodistas, filósofos, sicólogos, antropólogos, por mencionar algunos, y éstos serán acorde a los tipos, alcances y dimensiones de los proyectos donde se involucren. La convergencia misma de la ética con de tecnologías que involucran respuestas a veces indistinguibles de la creatividad humana, requiere de entendimientos y lenguajes más allá de lo tecnológico incluso.
\end{itemize}

\paragraph{DIMENSIONES COMPLEMENTARIAS}\mbox{} \\

El hecho de lo ético asumido a través del diseño orientado por valores como un eje central, es obligatorio en el debate sobre la inteligencia artificial, pero hay otras vías complementarias que también pueden abordar los riesgos asociados con la IA, ampliando la perspectiva y proporcionando soluciones holísticas. Estas vías incluyen:

\begin{itemize}
    \item \textbf{Regulación y legislación}: Más allá de los principios éticos, una regulación robusta y específica puede ser necesaria para establecer límites claros y ejecutables sobre cómo se puede desarrollar y utilizar la IA. Esto podría incluir legislación sobre privacidad de datos, uso de tecnologías de reconocimiento facial, y limitaciones en la automatización en ciertos sectores críticos.
\end{itemize}

\begin{itemize}
    \item \textbf{Diseño centrado en el humano}: Integrar enfoques de diseño que prioricen las necesidades, capacidades y limitaciones humanas. Esto asegura que los sistemas de IA se desarrollen de una manera que complemente las habilidades humanas y fomente la colaboración efectiva entre humanos y máquinas, en lugar de reemplazar o desplazar el trabajo humano.
\end{itemize}

\begin{itemize}
    \item \textbf{Trazabilidad e interpretabilidad}: Desarrollar y promover tecnologías que no solo sean eficientes sino también transparentes en sus procesos de toma de decisiones. Las inteligencias artificiales generativas permite recibir feedback sobre cómo se llega a los resultados y qué valores fueron influyentes, y este valor añadidos permite a los usuarios entender y confiar en cómo las decisiones son hechas por sistemas autónomos inteligentes e incide directamente en la aceptación y capacidad de supervisión efectiva de estos sistemas.
\end{itemize}

\begin{itemize}
    \item \textbf{Innovación abierta y colaborativa}: Fomentar modelos de desarrollo de IA que sean inclusivos y colaborativos, incluyendo a múltiples actores de diferentes sectores y disciplinas. La innovación abierta puede ayudar a diseminar el conocimiento y las mejores prácticas, y a asegurar que una gama más amplia de intereses y preocupaciones sean considerados durante el desarrollo de la tecnología.
\end{itemize}

\begin{itemize}
    \item \textbf{Desarrollo sostenible}: Alinear el desarrollo de la IA con los objetivos de desarrollo sostenible de la ONU y otros marcos éticos nacionales , regionales o locales para asegurar que contribuye tanto positivamente a desafíos globales como locales y a partir de aquí: a la reducción de la desigualdad,la mejora de la salud y el bienestar unido a la protección del medio ambiente de las comunidades donde está insertada la organización.
\end{itemize}

Estas vías nunca serán mutuamente excluyentes y, de hecho, serán ser más efectivas cuando se implementan de manera coordinada. Abordar los desafíos de la IA requiere un enfoque multifacético que combine la ética, la regulación, la innovación y la educación para crear un ecosistema tecnológico que sea seguro, justo y beneficioso para todos.

Por tanto, aunque podemos discutir y aplicar múltiples estrategias para abordar los desafíos específicos que presenta la IA, estas estrategias son efectivamente facetas de un compromiso más amplio con los valores. Cada uno de los enfoques que se han mencionado —desde la regulación hasta el diseño centrado en el humano y la trazabilidad — busca incorporar y reflejar estos valores éticos en prácticas concretas.

\begin{figure} [H]
    \centering
    \includegraphics[width=1.3\linewidth]{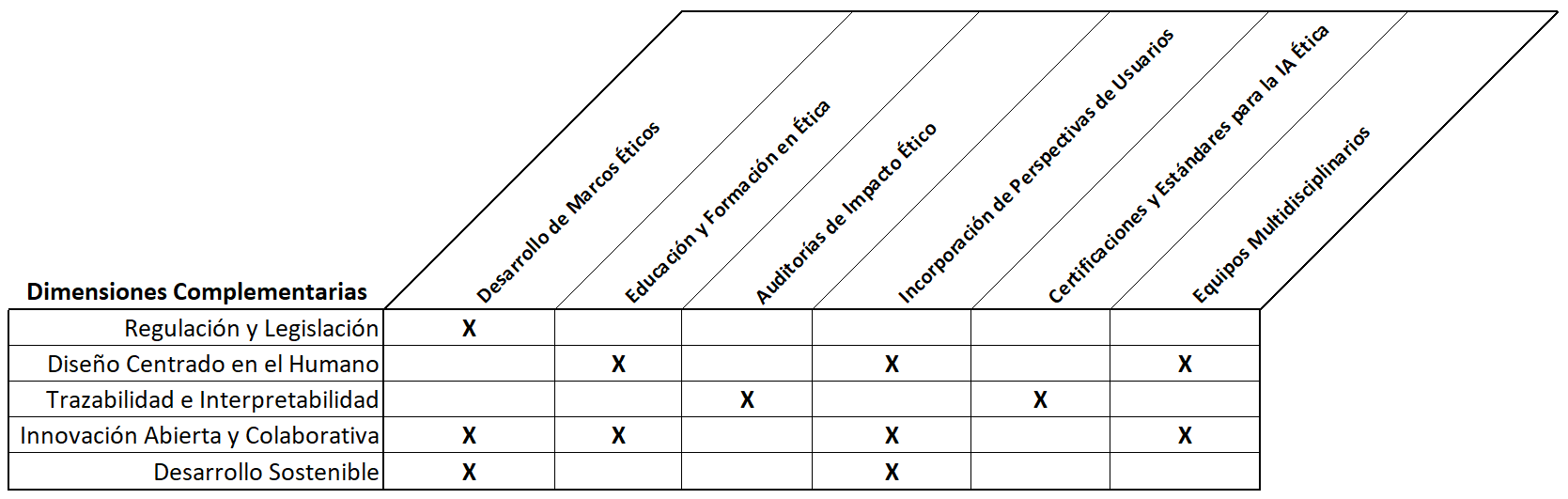}
    \caption{Potenciales intersecciones entre las dimensiones claves y las complementarias}
    \label{fig:claves-complementarias}
\end{figure}

Esta tabla (\textit{Figura 1}) ilustra la interacción entre las dimensiones claves y las complementarias. como consideraciones adicionales que refuerzan una implementación ética de la IA. Al usar estas conexiones, las organizaciones pueden identificar qué medidas complementarias son esenciales para respaldar cada dimensión clave, facilitando un enfoque holístico en la adopción de la IA.

\paragraph{ACCIONES ESPECÍFICAS}\mbox{} \\

Se requieren acciones específicas que las organizaciones puedan implementar para integrar principios éticos en el uso de la Inteligencia Artificial. Estas medidas concretas garantizan que el desarrollo y aplicación de la IA se alineen con valores éticos fundamentales, como la justicia y la transparencia, fortaleciendo así la responsabilidad y la confianza en las tecnologías emergentes.

\paragraph{\textbf{Acción: Definir principios éticos para la IA}}
\begin{itemize}
    \item Conformar un comité de ética que involucre a líderes, expertos y diversos grupos de interés para definir los principios éticos centrales que guiarán el uso de la IA en la organización.\cite{trejo2020hacia}
    \item Desarrollar y documentar formalmente un código de ética en IA que establezca principios como justicia, transparencia, responsabilidad y respeto a la privacidad como lineamientos obligatorios.
    \item Integrar este código de ética en la misión, valores y objetivos estratégicos de la empresa, asegurándose de que sea ampliamente comunicado y entendido por todos los niveles.
\end{itemize}

\paragraph{\textbf{Acción: Implementar prácticas de diseño ético de IA}}
\begin{itemize}
    \item Adoptar metodologías ágiles y centradas en el ser humano que involucren a empleados, clientes, comunidades y diversos stakeholders desde las primeras etapas del diseño de soluciones de IA.
    \item Incorporar evaluaciones de impacto ético y social como parte integral del proceso de diseño, identificando y mitigando riesgos potenciales antes del desarrollo e implementación.
    \item Formar equipos multidisciplinarios que incluyan perfiles técnicos, de negocio, ética, privacidad y representantes de grupos impactados para un enfoque integral.
\end{itemize}  

\paragraph{\textbf{Acción: Asegurar implementación responsable de IA}}
\begin{itemize}
    \item Establecer un marco de gobierno, políticas y procedimientos para monitorear continuamente el cumplimiento de los principios éticos durante todo el ciclo de vida de los sistemas de IA.
    \item Realizar auditorías éticas periódicas por entidades independientes para evaluar algoritmos, datos, procesos y resultados en busca de sesgos, riesgos o incumplimientos.
    \item Implementar soluciones técnicas de trazabilidad e interpretación para procesos que permitan comprender las razones detrás de resultados y decisiones que afectan directamente a personas.
\end{itemize}

\paragraph{\textbf{Acción: Realizar evaluaciones y ajustes continuos}}
\begin{itemize}    
    \item Crear canales accesibles (ej. líneas éticas, comunidades en línea) para que empleados, clientes y público puedan reportar inquietudes, sesgos observados u otros impactos éticos.
    \item Conformar comités de revisión que analicen estos reportes, investiguen a fondo, y recomienden acciones correctivas incluyendo ajustes a prácticas, procesos o sistemas de IA.
    \item Programar evaluaciones periódicas de impacto social y ético externas que permitan identificar oportunidades de mejora y tendencias que requieran cambios estratégicos.
\end{itemize}

\paragraph{\textbf{Acción: Implementar programas de educación y capacitación}}  
\begin{itemize}
    \item Desarrollar planes de formación continua obligatorios en todos los niveles, combinando conocimientos técnicos de IA con módulos de ética, sesgos, privacidad y derechos humanos.
    \item Incorporar actividades prácticas, casos de estudio y dilemas éticos que permitan a los empleados aplicar los principios aprendidos a situaciones reales.
    \item Instituir canales de comunicación interna, eventos y campañas que refuercen continuamente la importancia de la ética y el uso responsable de la IA para el bien común.
\end{itemize}

\begin{figure}[H]
    \centering
    \includegraphics[height=0.72\linewidth]{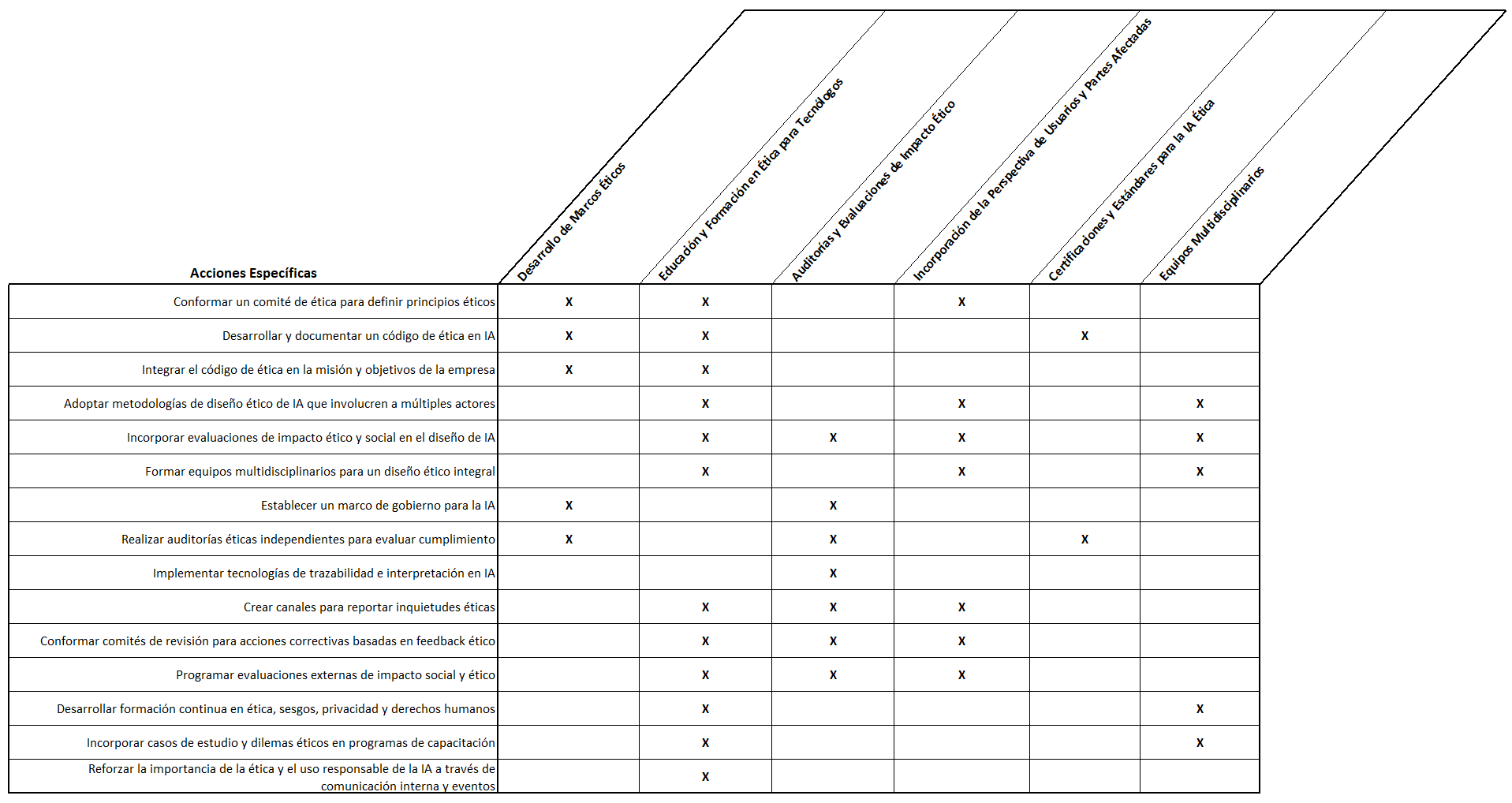}
    \caption{Acciones potenciales correspondientes a las dimensiones claves dentro de una estrategia de adopción de IA}
    \label{fig:dimensiones-acciones}
\end{figure}

La tabla de la figura 2 es una propuesta de qué acciones específicas pueden aplicarse para fortalecer las dimensiones clave en la implementación de la IA. Muestra cómo intervenciones concretas, como la definición de principios éticos o la implementación de prácticas de diseño ético, se alinean con áreas fundamentales como el desarrollo de marcos éticos, la educación en ética para tecnólogos, y las auditorías de impacto ético.

\paragraph{INCLUSION Y DIVERSIDAD}\mbox{} \\

Como un elemento aparte y distinto del contexto metodológico general, explicamos esta sección de "Género y Diversidad" para destacar la relevancia transversal y crítica de estos aspectos en la implementación de la inteligencia artificial. Estos temas no son simplemente adicionales ni secundarios; son fundamentales y deben integrarse en todas las dimensiones para garantizar la equidad, efectividad y justicia en la adopación o desarrollo de la IA.

 A medida que la IA se convierte en una herramienta más integral dentro de  una organización y sus procesos, es más crítico que nunca la necesidad de diseñar sistemas que reflejen y respeten la diversidad interna en especial, pero la de la sociedad en general. 
 
 Un componente muy conocido ya en los cuáles se puede incidir para regular de manera positiva son los algoritmos de entrenamiento de los modelos de IA, que a menudo son desarrollados y entrenados utilizando conjuntos de datos que no tienen en cuenta la diversidad de género, etnia, y otros factores sociodemográficos, lo que puede resultar en sesgos involuntarios que perpetúan la discriminación.
 
 Por ejemplo, los sistemas de reconocimiento facial han mostrado tasas de error significativamente más altas para mujeres y personas de color. Estos sesgos tecnológicos no son solo errores técnicos; reflejan y perpetúan desigualdades existentes en la sociedad. Veamos algunos sesgos conocidos y otros no tan obvios:
\begin{itemize}
    \item \textbf{Sesgo racial y étnico}: Algoritmos que no han sido adecuadamente ajustados pueden mostrar preferencias hacia ciertos grupos raciales o étnicos, debido a la representación desigual en los datos de entrenamiento. Por ejemplo: Las apps de citas en línea pueden ser propensas a sesgos raciales y reforzar estereotipos. Algunas de estas apps han sido criticadas por la discriminación racial presente en la elección de parejas.
    \item \textbf{Sesgo de edad}: Discriminación hacia individuos de ciertas edades, particularmente en decisiones de contratación y marketing, donde la IA podría favorecer a usuarios de ciertos rangos de edad, ignorando las necesidades y preferencias de otros grupos.
    \item \textbf{Sesgo de clase social y económica}: Decisiones automatizadas que favorecen a individuos de ciertos estratos socioeconómicos. Los algoritmos que evalúan el crédito pueden perpetuar la exclusión financiera de ciertos grupos, impidiendo su acceso a préstamos y oportunidades económicas. Por ejemplo, se descubrió que una importante compañía de tarjetas de crédito asignaba límites de crédito más bajos a las mujeres, incluso cuando sus perfiles financieros eran similares a los de los hombre
    \item \textbf{Sesgo lingüístico y cultural}: Algoritmos que no reconocen o valoran adecuadamente la diversidad lingüística y cultural, llevando a respuestas inadecuadas o irrelevantes para usuarios de distintas culturas o que hablan diferentes lenguas, especialmente minoritarias.
    \item \textbf{Sesgo geográfico}: La IA podría mostrar preferencias por usuarios de ciertas ubicaciones geográficas, basado en la prevalencia de datos de esos lugares en el conjunto de entrenamiento, descuidando así a usuarios en regiones menos representadas.
\end{itemize}

\paragraph{Estrategias para la promover la equidad}\mbox{}\\
Para combatir estos desafíos, las organizaciones deben adoptar estrategias específicas que aseguren la inclusión y la equidad en el desarrollo y uso de la IA \cite{unesco2023}. Algunas de esta estrategias son:

\begin{itemize}
    \item \textbf{Diversificación de Conjuntos de Datos:} Asegurarse de que los datos utilizados para entrenar algoritmos de IA sean representativos de la diversidad demográfica global. Esto incluye la recopilación de datos que sean suficientemente variados y la realización de pruebas rigurosas para identificar y corregir sesgos antes de que los algoritmos sean desplegados.
    
    \item \textbf{Equipos de Desarrollo Inclusivos:} Fomentar la diversidad dentro de los equipos de desarrollo de IA. Equipos diversos no solo mejoran la innovación y creatividad, sino que también pueden identificar potencialmente problemas de sesgo y equidad que podrían pasar desapercibidos en equipos más homogéneos.
    
    \item \textbf{Auditorías de Equidad Regular:} Implementar revisiones periódicas de los sistemas de IA para evaluar su desempeño y justicia. Las auditorías deben ser realizadas tanto internamente como por terceros independientes para garantizar la transparencia y la objetividad.
    
    \item \textbf{Educación y Sensibilización:} Capacitar a todos los empleados sobre la importancia de la diversidad y la inclusión en la tecnología. Esto incluye formación específica sobre cómo los sesgos pueden infiltrarse en los algoritmos y las maneras de mitigar estos riesgos en el diseño y la implementación de sistemas de IA.
    
    \item \textbf{Participación Comunitaria:} Involucrar a comunidades diversas en el proceso de desarrollo de IA. Obtener retroalimentación de una amplia gama de usuarios puede ayudar a identificar problemas de equidad y mejorar la aceptación y eficacia de los sistemas de IA.
    
\end{itemize}
\section{Cuatro pilares para la implementación de un marco ético en las organizaciones}

Si bien comprender las diversas dimensiones, actores o acciones presentadas es fundamental para abordar de manera holística los desafíos éticos de la inteligencia artificial, es igualmente determinante contar con un enfoque metodológico estructurado. Este marco permitirá a las organizaciones integrar coherentemente los principios éticos en cada etapa del ciclo de vida de los sistemas de IA, desde la estrategia hasta las operaciones diarias.

En esta línea, se proponen cuatro pilares fundamentales que actuarán como guías rectoras para que cualquier organización pueda construir un abordaje robusto e integral orientado a la incorporación ética de la Inteligencia Artificial. Estos pilares brindan una visión metódica y unificada, aglutinando los diversos componentes éticos, sociales y de gobernanza analizados previamente en un marco práctico y accionable para las organizaciones.

\begin{figure}[H]
    \centering
    \includegraphics[width=1\linewidth]{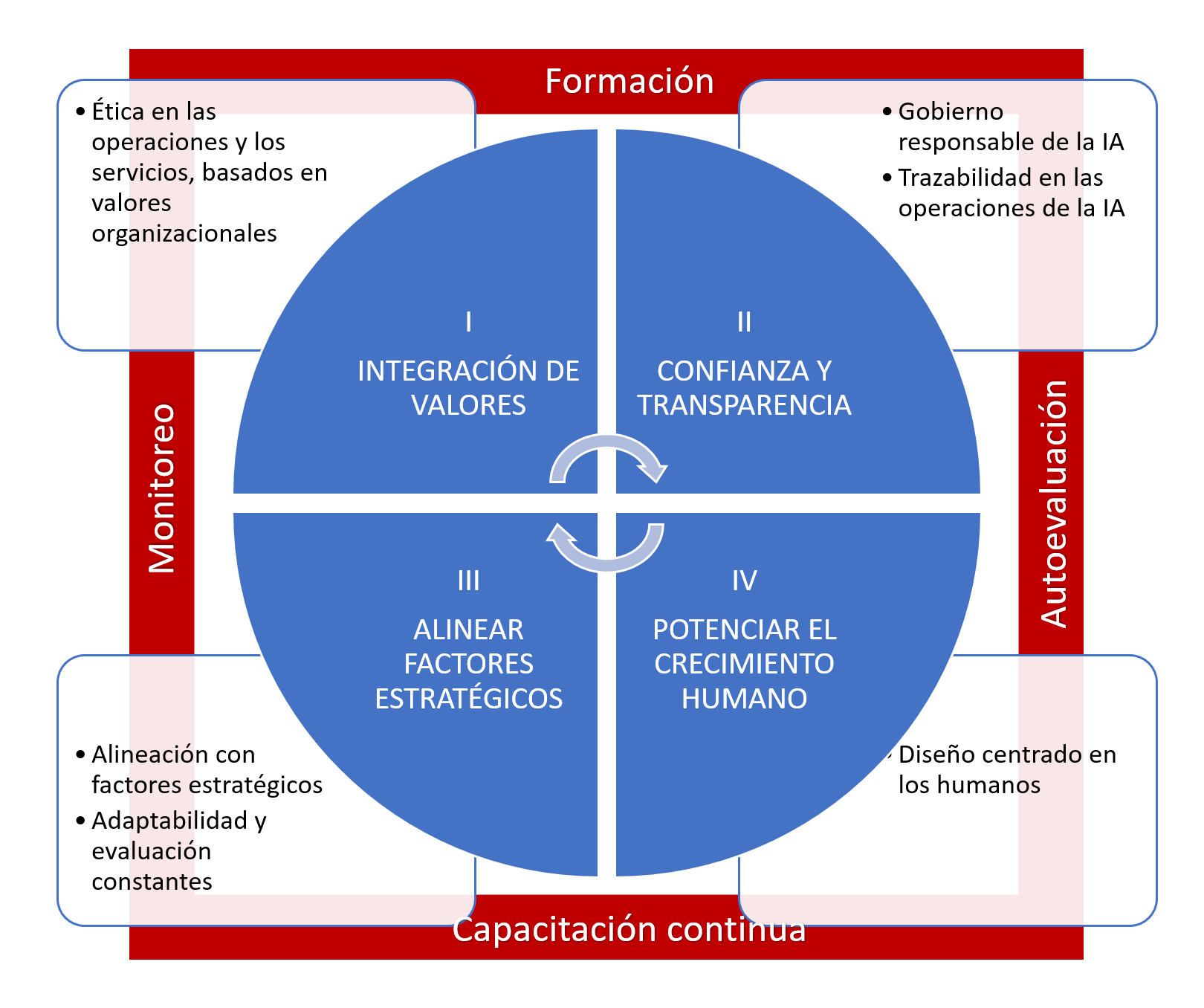}
    \caption{\textit{Pilares para un marco ético para la adopción y desarrollo de la Inteligencia Artificial en las organizaciones}}
\end{figure}

\paragraph{PILAR I: INTEGRACIÓN DE VALORES:}\mbox{}\\

Este pilar se centra en incrustar los valores de la empresa en todos los aspectos del desarrollo y uso de la IA, asegurando que las tecnologías reflejen y promuevan la identidad y los estándares éticos de la organización. La formación continua es es un requisito crítico para mantener a todos los empleados alineados con estos valores. No se trata simplemente de establecer principios éticos en documentos, sino de lograr que estos valores se vuelvan intrínsecos a la cultura organizacional y se reflejen en cada etapa del desarrollo y uso de la IA.

La clave radica en involucrar a todos los empleados, desde el liderazgo hasta los equipos técnicos y operativos, en un proceso de revisión y alineación constantes con la identidad ética de la empresa. Esto implica programas de formación continua que no solo transmitan conocimientos técnicos, sino también un profundo entendimiento de cómo aplicar los valores organizacionales en la toma de decisiones y el diseño de soluciones de IA.

Además, es fundamental establecer procesos y políticas que traduzcan estos valores en prácticas concretas. Por ejemplo, auditorías de algoritmos para detectar y mitigar sesgos, protocolos de privacidad de datos, o mecanismos de rendición de cuentas ante posibles impactos negativos. De esta manera, los valores dejan de ser meras declaraciones y se convierten en lineamientos operacionales que guían el trabajo diario de todos los involucrados en el desarrollo de la IA.

\paragraph{PILAR II: CONFIANZA Y TRANSPARENCIA:}\mbox{}\\

Enfatiza la gobernanza de la IA, transparencia y responsabilidad. Los "embajadores de IA" son clave para supervisar y asegurar la calidad y diversidad de los datos, así como para proporcionar un feedback continuo, mejorando así la confiabilidad y equidad de los sistemas de IA.

Se han de implementar estructuras de gobernanza sólidas, que incluyan comités de ética, auditorías independientes y mecanismos de supervisión continua. Estos organismos deben velar por el cumplimiento de los principios éticos y garantizar que los sistemas de IA operen de manera justa, transparente y responsable.

Otro aspecto imprescindible es el desarrollo de sistemas de IA interpretables y con trazabilidad, que permitan a los usuarios comprender las razones detrás de las decisiones tomadas y los factores que influyen en los resultados. Esto no solo promueve la confianza en los sistemas de IA, sino que también facilita la identificación y corrección de posibles sesgos o errores al poder rastrear los datos, reglas y procesos que llevaron a determinados resultados. 

Finalmente, es fundamental establecer canales de comunicación abiertos y bidireccionales con todas las partes interesadas, incluyendo empleados, clientes y comunidades afectadas. Esto permite recibir retroalimentación valiosa, abordar preocupaciones y mantener un diálogo constante sobre el impacto de la IA en la sociedad

\paragraph{PILAR III: POTENCIAR EL CRECIMIENTO HUMANO:}\mbox{}\\

Al enfocarse en la colaboración entre la inteligencia humana y artificial, la implementación de este pilar busca potenciar la creatividad y minimizar las tareas monótonas, destacando la importancia de adaptarse a nuevos roles y continuo aprendizaje. El objetivo es aprovechar las fortalezas únicas de la IA para potenciar las capacidades humanas, liberando a las personas de tareas repetitivas y permitiéndoles centrarse en actividades más creativas y gratificantes.

El diseño centrado en el humano es una decisión clave que implica la participación activa de usuarios y empleados en el desarrollo de soluciones de IA. Esto garantiza que los sistemas se adapten a las necesidades y limitaciones humanas, en lugar de forzar a las personas a adaptarse a la tecnología.

Además, es fundamental invertir en programas de capacitación y desarrollo continuo para preparar a la fuerza laboral para los cambios impulsados por la IA. Esto incluye el desarrollo de habilidades complementarias, como el pensamiento crítico, la resolución de problemas y la comunicación efectiva, que serán esenciales en un entorno de trabajo cada vez más automatizado.

\paragraph{PILAR IV: ALINEAR FACTORES ESTRATÉGICOS:} \mbox{}\\

Si bien la ética y los valores deben ser la base de cualquier iniciativa de IA, este pilar reconoce la importancia de alinear estas tecnologías con los objetivos estratégicos de la organización. Se trata de identificar y aprovechar los factores clave que pueden potenciar el uso de la IA para impulsar el crecimiento, la eficiencia y la competitividad de la empresa.

Esto requiere un enfoque adaptable y práctico, que implique un monitoreo constante de las tendencias tecnológicas y del mercado. Las organizaciones deben estar preparadas para ajustar sus estrategias y adoptar nuevas soluciones de IA a medida que surjan oportunidades o desafíos.

Y es fundamental realizar evaluaciones exhaustivas de los riesgos y beneficios potenciales de cada implementación de IA. Esto implica considerar no solo los impactos financieros, sino también los aspectos éticos, sociales y ambientales, asegurando que la innovación se lleve a cabo de manera responsable y sostenible.

\section{Conclusiones}
La implementación ética de la inteligencia artificial en las organizaciones es más que una responsabilidad; es una estrategia esencial que debe preceder incluso a la definición de objetivos de negocio. El establecimiento de marcos éticos rigurosos no solo guía el desarrollo y uso de la IA, sino que también protege a la organización contra riesgos futuros y refuerza su reputación a largo plazo. Estos marcos no son simplemente un complemento a las estrategias tecnológicas y comerciales, sino fundamentos sobre los cuales se construyen.

Para las organizaciones de hoy, la tarea de integrar la ética en la IA implica varias acciones clave. Primero, es importante que la alta dirección adopte y promueva una visión ética que se integre en todas las fases del ciclo de vida de la IA, desde la conceptualización hasta el despliegue y la revisión operativa. Esto significa que las decisiones sobre IA deben ser revisadas no solo por su impacto económico, sino también por su alineación con principios éticos como la transparencia, la justicia, y el respeto por la privacidad y la dignidad humana.

En segundo lugar, las organizaciones deben ser proactivas en la formación de sus equipos, no solo en habilidades técnicas, sino también en competencias éticas. Esto incluye la comprensión de las implicaciones sociales de la IA, la capacidad para identificar potenciales sesgos en algoritmos y datos, y el desarrollo de soluciones que respeten los derechos y valores de todos los stakeholders. La formación en ética debe ser una parte continua de la educación corporativa, adaptándose a las nuevas tendencias y desafíos que surjan con la evolución de la tecnología.

Además, es fundamental la implementación de procesos de auditoría y evaluación continua que aseguren que los sistemas de IA operen dentro de los límites éticos establecidos. Estos procesos deben incluir no solo revisiones internas, sino también la participación de terceros independientes que puedan ofrecer nuevas perspectivas y ayudar a evitar la "ceguera organizativa" respecto a los problemas éticos.

Por último, la transparencia con los consumidores y la sociedad en general es esencial. Las organizaciones deben comunicar abierta y honestamente cómo utilizan la IA, los datos que recopilan y los controles que tienen en lugar para proteger esos datos y asegurar un trato justo y equitativo. Esto no solo es una práctica ética, sino que también contribuye a construir confianza, un activo invaluable en la era digital.

En general la ética debe ser la piedra angular en la estrategia de cualquier organización que aspire a incorporar las nuevas tecnologías de inteligencia artificial. Antes de perseguir el potencial de negocio que la IA puede ofrecer, es imperativo establecer un marco sólido que garantice que la tecnología se desarrolle y utilice de manera que respete y promueva los valores humanos. Al hacerlo, las organizaciones no solo aseguran su propio futuro sostenible, sino que también contribuyen al bienestar de la sociedad en su conjunto.

\newpage

\printbibliography[
heading=bibintoc,
title={Bibliografía}
]

\end{document}